\documentclass[aps,prd,preprintnumbers,nofootinbib,onecolumn]{revtex4}
\usepackage{bm}
\usepackage{latexsym}
\usepackage{dcolumn}
\usepackage{amsmath,amsfonts,amssymb}
\usepackage{graphicx,epsfig}
\usepackage{color}
\usepackage{amsthm}

\begin{document}
\author{Mir  Hameeda$^{a,b}$, Mir Faizal$^c$, Ahmed Farag Ali$^{d,e}$\\
$^a$Department of Physics
S.P. Collage,  Srinagar, Kashmir, 190001 India
\\
$^b$Visiting Associate, IUCCA,  Pune,  41100, India
\\
$^c$Department of Physics and Astronomy,   University of Waterloo, Waterloo,
ON N2L 3G1, Canada
\\
$^d$Department of Physics,
Florida State University,
Tallahassee   FL 32306 USA
\\
$^e$Department  of Physics, Faculty of Science,
 Benha University, Benha, 13518, Egypt}
\title{Clustering of Galaxies in Brane World Models   }

\begin{abstract}
In this paper, we analyze the clustering of galaxies using a modified Newtonian potential.
This modification of the Newtonian potential occurs due to the existence  of extra dimensions in brane world models.  We will analyze
a system of galaxies interacting with each other through this modified Newtonian potential.
The partition function for this system of galaxies will be calculated, and this partition function will be used
to calculate the free energy of this system of galaxies. The entropy and the chemical potential for this system will
also be calculated. We will derive explicit expression for the clustering parameter for this system.
This parameter will determine the behavior of this system, and we will be able to express various thermodynamic
quantities using this clustering parameter. Thus, we will be able to explicitly analyze the effect that modifying
the Newtonian potential can have on the clustering of galaxies. We also analyze the effect of extra dimensions on the two-point functions between
galaxies.
\end{abstract}

\maketitle

\section{Introduction}
Motivated by the existence of extra dimensions in string theory, brane world models propose
that our universe is a brane in a higher dimensional bulk
\cite{Randall:1999vf,Randall:1999ee}. The extra dimensions can be either compactified or infinite,
and the number of such extra dimensions along with  the field content of the theory depends on the details of
the model that is being investigated.
    However,  common feature of  all of these
different brane models is that the standard model fields are confined to the four dimensional brane.
The gravitons are not confined to the brane and can propagate into the higher dimensional bulk.
This explains the weakness of the
gravitational field \cite{Randall:1999vf,Randall:1999ee,ArkaniHamed:1999dc,ArkaniHamed:1998nn,
Antoniadis:1998ig,ArkaniHamed:1998rs}. In fact, it is hoped that these models can also provide
a solution for the solution of the hierarchy problem \cite{Randall:1999vf,Randall:1999ee,
ArkaniHamed:1999dc,ArkaniHamed:1998nn,Antoniadis:1998ig,ArkaniHamed:1998rs}.
It may be noted
that the Newtonian potential gets modified at large distances due to the propagation of gravity
into the bulk \cite{Floratos:1999bv}. These corrections are parametrized by Yukawa potential
\cite{Long:2003dx}. Thus, because of the existence of extra dimensions,
the original Newtonian potential is modified to \cite{Long:2003dx}
\begin{equation}
 \Phi  = \Phi_N (1 + \alpha e^{\lambda r}),
\end{equation}
where $\alpha, \lambda$ depend on the details of model, and $V_N$ is the original Newtonian potential,
\begin{equation}
 \Phi_N = - \frac{G m^2 }{r}.
\end{equation}
However, in the Randall-Sundrum model the modification to the Newtonian potential
is parameterized by a power law deformation of the usual Newtonian potential \cite{Randall:1999vf,Randall:1999ee}
\begin{equation}
 \Phi  = \Phi_N \left( 1 + \frac{k}{r}\right).
\end{equation}
The action for the Randall-Sundrum model is given by \cite{a}
\begin{equation}
 S = S_{bulk} + S_{brane},
\end{equation}
where
\begin{eqnarray}
 S_{bulk} &=& \int d^4x dy \sqrt{|g_5|} (  M R_5 -
 \Lambda_5),  \\
 S_{brane}&=& \int d^4x \sqrt{|g_4|} V_{brane}.
\end{eqnarray}
Here $R_5$ is the five dimensional scalar curvature, $g_5$
and $g_4$ are determinants of the five dimensional and the
four dimensional metrics, respectively. The $M$ denotes the
a suitable constant formed from the
five dimensional Planck mass. The brane tension is denoted by
$V_{brane}$. The four dimensional metric is induced on the
hyper-surface $y =0$. In the weak field approximation, this model gives a power law deformation of the usual Newtonian potential \cite{a}.

Unlike many other brane world models, the Randall-Sundrum model   allows for the existence of infinite extra
dimensions. It is this power law deformation of the usual Newtonian potential which we will use for our analysis
in this paper. It is worth mentioning that there are different approaches where    Newton's law of gravity get
modified. One of these approaches is the non-commutative geometry
\cite{Nicolini:2010nb,Gregory:2012an}. Besides, some modifications of Newton's law of
gravity have been found due to minimal length in quantum gravity \cite{Ali:2013ma}. In addition,
a modification of Newton's law of gravity has been predicted in entropic gravity \cite{Setare:2010ct}.
Besides, another sources of modifications to Newton's law of gravity have been investigates in
f(R) theories \cite{Nojiri:2007cq}. It may be noted that even a modification   based on entropic force produces similar results \cite{barun, 10}.
Even though it is related some minimal length deformation, the important thing to note here is that this modification of Newtonian potential can be motivated from various physical phenomena.
So, in this paper, we will use the modified  Newtonian potential based on Randall-Sundrum models for analyzing the clustering of galaxies.

The modification of the Newtonian potential can occur only at very small or very large distances,
at which the Newtonian theory of gravity is not tested. In fact, it has been observed that the rotation
curves of galaxies do not fit the theoretically predicted curves, and this has motivated the development
of theories like the MOND \cite{MOND, MOND2} and MOG \cite{MOG, MOG2}. So, we expect the deviation from the
Newtonian potential to become effective at galactic scale. In this paper, we will study a system
of   of galaxies interacting gravitationally with each other. This system  is large  enough
the modification to the Newtonian potential to become significant.
We will analyse the effect this modification of the
Newtonian potential has on the  clustering of galaxies. In our analysis, we will effectively treat
each galaxy as a point particle. However, we will incorporate an random parameter, which will take into account
the fact that galaxies do actually have an intended structure.
As we will be dealing with a large number of galaxies, and individual galaxies will be effectively treated
as point particles, we can apply the methods of statistical mechanics and calculate the partition function
for this system \cite{sas85, ahm10, sas84, ahm02, ahm06}. The strength of interaction will depend effectively on the
average density of the universe  \cite{sas85}.

A critical  physical assumption in this analysis is that the  local dynamical timescale in an over
dense region  varies    faster than global gravitational time scale.
This assumption makes it possible for gravitational clustering to evolve through a sequence of
quasi-equilibrium state \cite{ahm10}.
In fact, using this assumption the
cosmological many body partition function has been calculated  \cite{ahm02, ahm14, ahm144}. This was done by
using  using an ensemble of co-moving cells, and each of these co-moving cells contained
   galaxies which interacted with each other. It was demonstrated that if the size of the
cells is smaller than the particle correlation length, then each member of this ensemble is correlated
gravitationally with other cells. Furthermore, the correlations within  each cells is greater than correlations among
different cells, so that extensive is justified as a  good approximation \cite{sas00,  ahm02}. In this paper,
we will  use this formalism for analyzing a system of galaxies interacting through a modified Newtonian
potential.

\section{Gravitational Partition Function }
In this section, we will calculate the partition function for a system of galaxies interacting though a modified Newtonian potential.
 We consider a large system, which consists of an ensemble of cells, all of the
same volume $V$, or radius $R_{1}$ and average density $\bar\rho$. Both the number of galaxies and their total energy will vary among
these cells, thus these are represented by a grand canonical ensemble. Now the  galaxies within the system have gravitational pairwise interaction
generated by the modified Newtonian potential.
It will be further assumed that the distribution is statistically homogeneous over large regions.

The general partition function of a system of $N$ particles of mass $m$ interacting through the modified gravitational potential
with potential energy is $\Phi$,  can be written as
\begin{eqnarray}
Z(T,V)&=& \frac{1}{\Lambda^{3N}N!}\int d^{3N}pd^{3N}r
\times  exp\biggl(-\biggl[\sum_{i=1}^{N}\frac{p_{i}^2}{2m}+\Phi(r_{1}, r_{2}, r_{3},
\dots, r_{N})\biggr] T^{-1}\biggr),
\end{eqnarray}
where $p_{i}$ is the momentum for different galaxies  and $T$ is the average temperature.
It may be noted that a similar analysis has been done for galaxies interacting though the usual Newtonian potential \cite{ahm02}.
Here $N!$ takes the distinguish-ability
of classical particles into account, and $\Lambda$ is the normalization factor which results from integration over momentum space.
Now integrating the   momentum space, we obtain the following expression
\begin{eqnarray}
Z_N(T,V)=\frac{1}{N!}\big(\frac{2\pi mT}{\Lambda^2}\big)^{3N/2}Q_N(T,V),
\end{eqnarray}
where $Q_{N}(T,V)$ is the configuration integral
\begin{eqnarray}
Q_{N}(T,V)=\int....\int \prod_{1\le i<j\le N} exp[-\phi_{ij}(r_{1}, r_{2},\dots,r_{N})T^{-1}]d^{3N}r.
\end{eqnarray}
The gravitational potential energy $\Phi(r_{1}, r_{2}, \dots, r_{N})$ is
a function of the relative position vector $r_{ij}=|r_{i}-r_{j}|$
and is the sum of the potential energies of all pairs.
Even for a  system of galaxies in brane world models,
the potential energy $\Phi(r_{1}, r_{2}, \dots, r_{N})$ can be expressed as
\begin{eqnarray}
\Phi(r_{1}, r_{2}, \dots, r_{N})=\sum_{1\le i<j\le N}\Phi(r_{ij})=\sum_{1\le i<j\le N}\Phi(r).
\end{eqnarray}
In fact, we can use  a two particle function
$f_{ij}=e^{-\Phi_{ij}/T}-1$, such that it
vanishes  in absence of interactions,  and  is non-zero only for interacting galaxies.

It has been demonstrated that for galaxies interacting through a usual Newtonian potential,
that the configurational integral can be expressed in terms of the function $f_{ij}$ \cite{ahm02}.
Repeating this analysis for galaxies interacting through modified Newtonian potential, we obtain
\begin{eqnarray}
Q_{N}(T,V)&=& \int....\int d^{3}r_{1}d^{3}r_{2}\dots d^{3}r_{N}
\biggl[(1+f_{12})(1+f_{13})(1+f_{23})(1+f_{14})\dots (1+f_{N-1,N})\biggr].
\end{eqnarray}
The interaction potential energy between galaxies in brane world model with large extra dimensions,
can be written as \cite{Randall:1999vf,Randall:1999ee}
\begin{eqnarray}
\phi_{i,j}=-\frac{Gm^2}{ r_{ij}    }\biggl(1+\frac{k_l}{ r_{ij}    }\biggr),
\end{eqnarray}
where $r>>{|k_l|}^{l/2}$ and ${|k_l|}^{l/2}$ is considered as a
typical length scale at which the correction due to the infinite extra dimensions become dominant.
It may be noted that for  point masses  the partition function  for galaxies interacting through usual Newtonian potential
 diverges at $r_{ij}=0$.
This divergence occurs due to the assumption that the galaxies are point like objects.
  However,
  this  divergence can be    removed by taking the extended nature of galaxies into account by introducing
a softening parameter which takes care of the finite size of each galaxy \cite{ahm10, ahm02, ahm06}. Thus by incorporating the softening parameter
the interaction potential energy between galaxies in brane world models can be represented as
\begin{eqnarray}
\phi_{i,j}=-\frac{Gm^2}{(r_{ij}^2+\epsilon^2)^{1/2}}\biggl(1+\frac{k_l}{(r_{ij}^2+\epsilon^2)^{l/2}}\biggr),
\end{eqnarray}
where $\epsilon$ the  softening parameter which is related to the extended nature of galaxies.
Now we can write
\begin{eqnarray}
f_{i,j}=exp{\biggl[\frac{Gm^2}{(r_{ij}^2+\epsilon^2)^{1/2}}\biggl(1+\frac{k_l}{(r_{ij}^2+\epsilon^2)^{l/2}}\biggr)\biggr]}-1.
\end{eqnarray}
Expanding the   function $f_{ij}$, we obtain the following expression.
\begin{eqnarray}
f_{ij}&=& \biggl[\frac{Gm^2}{(r_{ij}^2+\epsilon^2)^{1/2}}\biggl(1+\frac{k_l}{(r_{ij}^2+\epsilon^2)^{l/2}}\biggr)\biggr] +
\frac{1}{2!}\biggl[\frac{Gm^2}{(r_{ij}^2+\epsilon^2)^{1/2}}\biggl(1+\frac{k_l}{(r_{ij}^2+\epsilon^2)^{l/2}}\biggr)\biggr]^{2}\nonumber\\ && +
\frac{1}{3!}\biggl[\frac{Gm^2}{(r_{ij}^2+\epsilon^2)^{1/2}}\biggl(1+\frac{k_l}{(r_{ij}^2+\epsilon^2)^{l/2}}\biggr)
\biggr]^{3}
+\frac{1}{4!}\biggl[\frac{Gm^2}{(r_{ij}^2+\epsilon^2)^{1/2}}\biggl(1+\frac{k_l}{(r_{ij}^2+\epsilon^2)^{l/2}}\biggr)\biggr]^{4} +\dots.
\end{eqnarray}
Now we obtain $Q_{1}(T,V)=V$ by evaluating the configuration integrals over a spherical volume of radius $R_{1}$.  Similarly,
evaluating the configuration integrals  for
  $Q_{2}(T,V)$, we obtain
\begin{eqnarray}
Q_{2}(T,V)&=& 4\pi V\sum_{n=0}^{\infty}\frac{1}{n!}\big(\frac{Gm^{2}}{T}\big)^{n}\int_{0}^{R_{1}}\frac{r^{2}}{(r^{2}+\epsilon^{2})^{n/2}}
\times \biggl(1+\frac{k_l}{(r_{ij}^2+\epsilon^2)^{l/2}}\biggr)^ndr.
\end{eqnarray}
For large values of $r$, we can neglect  the higher terms  corresponding to   $\frac{k_l}{(r_{ij}^2+\epsilon^2)^{l/2}}$, and thus, we can write
\begin{eqnarray}
Q_{2}(T,V)&=& 4\pi V\sum_{n=0}^{\infty}\frac{1}{n!}\big(\frac{Gm^{2}}{T}\big)^{n} \times\int_{0}^{R_{1}}\frac{r^{2}}{(r^{2}+\epsilon^{2})^{n/2}}
\biggl(1+n\frac{k_l}{(r^2+\epsilon^2)^{l/2}}\biggr)dr
\nonumber \\
&=&4\pi V\sum_{n=0}^{\infty}\frac{1}{n!}
\big(\frac{Gm^{2}}{T}\big)^{n} \times\int_{0}^{R_{1}}\biggl(\frac{r^{2}}{(r^{2}+\epsilon^{2})^{n/2}}+n\frac{k_l}{(r^2+\epsilon^2)^{{(l+n)}/2}}\biggr)dr.
\end{eqnarray}
Evaluating this  integrals,  we obtain
\begin{eqnarray}
Q_{2}(T,V)=V^2\big(1+\sum_{n=1}^{\infty}(\alpha_{n}x^{n}+\beta_{n}x^n)\big),
\end{eqnarray}
where $x=3/2\beta \bar\rho T^{-3}$. The values of $\alpha_n$ and $\beta_n$ are given by
\begin{eqnarray}
\alpha_n&=& \frac{1}{n!}\biggl(\frac{2R_{1}}{3\epsilon}\biggr)^{n} F \biggl(\frac{3}{2},\frac{n}{2};\frac{5}{2};-\frac{R_{1}^2}{\epsilon^2}\biggr),
\nonumber \\
\beta_n&=& n\frac{k_l}{\epsilon^n}\frac{1}{n!}\biggl(\frac{2R_{1}}{3\epsilon}\biggr)^{n} F \biggl(\frac{3}{2},\frac{(n+l)}{2};\frac{5}{2};
-\frac{R_{1}^2}{\epsilon^2}\biggr),
\end{eqnarray}
where  $ F \biggl(\frac{3}{2},\frac{n}{2};\frac{5}{2};-\frac{R_{1}^2}{\epsilon^2}\biggr)$ is the hyper-geometric function.
Now as $R_{1}\sim \rho^{-1/3}$, we can write
\begin{eqnarray}
\frac{3Gm^{2}}{2R_{1}T}=\frac{3Gm^{2}}{2\rho^{-1/3}T}=\frac{3}{2}Gm^{2}\rho^{1/3}T^{-1}.
\end{eqnarray}
Using scale invariance,  $\rho\to\lambda^{-3}\rho$, $T\to\lambda^{-1}T$ and $r\to\lambda r$, we  obtain
\begin{eqnarray}
\frac{3}{2}(Gm^2)^{3}\rho T^{-3}=\beta\rho T^{-3},
\end{eqnarray}
where $\beta= \frac{3}{2}(Gm^{2})^{3} $.
Thus,  we can write
\begin{eqnarray}
Q_{2}(T,V)=V^2\big(1+\sum_{n=1}^{\infty}(\alpha_{n}+\beta_{n})x^{n}\big),
\end{eqnarray}
where $x=\beta \bar\rho T^{-3}$.
By similar procedure, we obtain configuration integral  an expression for general term as
\begin{eqnarray}
Q_{N}(T,V)=V^N\big(1+\sum_{n=1}^{\infty}(\alpha_{n}+\beta_{n})x^{n}\big)^{N-1}.
\end{eqnarray}
Thus, we can write
\begin{eqnarray}
Q_{3}(T,V)&=& V^3\big(1+\sum_{n=1}^{\infty}(\alpha_{n}+\beta_{n})x^{n}\big)^{2},
 \\
Q_{4}(T,V)&=& V^4\big(1+\sum_{n=1}^{\infty}(\alpha_{n}+\beta_{n})x^{n}\big)^{3}.
\end{eqnarray}

Hence,  the gravitational partition function for a system of galaxies in the brane world model, can be written as
\begin{eqnarray}
Z_N(T,V)=\frac{1}{N!}\big(\frac{2\pi mT}{\Lambda^2}\big)^{3N/2}V^{N}\big(1+\sum_{n=1}^{\infty}(\alpha_{n}+\beta_{n})x^{n}\big)^{N-1}
\end{eqnarray}
It may be noted that the weakly interacting  system can be represented by $n=1$, and the
 strongly interacting systems can be represented by taking higher values of $n$.
\section{Thermodynamics of Galaxies }
Now we will use the partition function for the system of galaxies interacting in a brane world model to calculate relevant thermodynamic quantities.
We can define    $B_{n}$ for different values of $n$ as
\begin{eqnarray}
B_{n}=\frac{\sum_{n=1}^{\infty} n(\alpha_{n}+\beta_{n})x^{n}}{1+\sum_{n=1} (\alpha_{n}+\beta_{n})x^{n}}.
\end{eqnarray}
This is the general clustering parameter for gravitationally interacting system of galaxies in brane world models.
We can express various thermodynamic quantities in terms of this general clustering parameter.
Thus, we can  write the
Helmholtz free energy $F=-T\ln Z_{N}(T,V)$ as
\begin{eqnarray}
F=  -T\ln \biggl(\frac{1}{N!}\big(\frac{2\pi mT}{\Lambda^2}\big)^{3N/2}V^N\big(1+\sum_{n=1}^{\infty}(\alpha_{n}+\beta_{n})x^n\big)^{N-1}\biggr).
\end{eqnarray}
The entropy $S$ can now be calculated from this Helmholtz free energy,
\begin{eqnarray}
S&=& -\biggl(\frac{\partial F}{\partial T}\biggr)_{N,V}
 \nonumber \\
 &=& N\ln (\rho^{-1}T^{3/2})+(N-1)\ln \big(1+\sum_{n=1}^{\infty}(\alpha_{n}+\beta_{n})x^{n}\big) \nonumber \\ &&  -3N
 \frac{\sum_{n=1}^{\infty} n(\alpha_{n}+\beta_{n})x^{n}}{1+\sum_{n=1}^{\infty} (\alpha_{n}+\beta_{n})x^n}
 +\frac{5}{2}N+\frac{3}{2}N\ln \big(\frac{2\pi m}{\Lambda^2}\big).
\end{eqnarray}
Now using the expression, for $B_n$
and for large $N$, and also using $N-1\approx N$, we obtain
\begin{eqnarray}
\frac{S}{N}=\ln (\rho^{-1}T^{3/2})+\ln \big(1+\sum_{n=1}^{\infty}(\alpha_{n}+\beta_{n})x^{n}\big)-3B_{n}+\frac{S_{0}}{N},
\end{eqnarray}
where $S_{0}=\frac{5}{2}N+\frac{3}{2}N\ln \big(\frac{2\pi m}{\Lambda^2}\big)$.
The internal energy $U =  F+TS$ and  of a system of galaxies can now expressed as
\begin{eqnarray}
U = \frac{3}{2}NT\big(1-2B_{n}\big).
\end{eqnarray}
Similarly, we can write the  pressure $P$   and chemical potential $\mu$ can be expressed as
\begin{eqnarray}
P&=& -\biggl(\frac{\partial F}{\partial V}\biggr)_{N,T}
=\frac{NT}{V}\big(1-B_{n}\big),
\\
\mu &=& \biggl(\frac{\partial F}{\partial N}\biggr)_{V,T}
 =  T \ln (\rho T^{-3/2})- T \ln \big(1+\sum_{n}^{\infty}(\alpha_{n}+\beta_{n})x^{n}\big)- T \frac{3}{2}\ln \big(\frac{2\pi m}{\Lambda^2}\big)- T B_{n},.
\end{eqnarray}

 The probability of finding $N$ particles can be written as
\begin{eqnarray}
F(N)=\frac{\sum_{i}e^{\frac{N\mu}{T}}e^{\frac{-U_{n}}{T}}}{Z_{G}(T,V,z)}=\frac{e^{\frac{N\mu}{T}}Z_{N}(V,T)}{Z_{G}(T,V,z)},
\end{eqnarray}
where $Z_{G}$ is the grand partition function defined by
\begin{eqnarray}
Z_{G}(T,V,z)=\sum_{N=0}^{\infty}z^NZ_{N}(V,T).
\end{eqnarray}
and $z$ is the activity.
Thus, for a system of gravitationally interacting system of galaxies in brane world models, we can write
\begin{eqnarray}
e^{\frac{N\mu}{T}}=\biggl(\frac{{\bar N}}{V}T^{-3/2}\biggr)^{N}\biggl(1+\sum_{n=1}^{\infty}\frac{(\alpha_{n}+\beta_{n})}{(\alpha_{1}+\beta_{1})^{n}}\frac{B^n}{(1-B)^{n}}\biggr)^{-N}
e^{-NB_{n}}\biggl(\frac{2\pi m}{\Lambda^{2}}\biggr)^{-3N/2},
\end{eqnarray}
Now the  grand partition function for this  system of gravitationally interacting system of galaxies in brane world models,can be written as
\begin{eqnarray}
\ln Z_{G}=\frac{PV}{T}=\bar N(1-B_{n}),
\end{eqnarray}
and the distribution function can be expressed as
\begin{eqnarray}
F(N)_n &=& \frac{\bar {N}^{N}}{N!}\biggl(1+\sum_{n=1}^{\infty}\frac{N^{n}(\alpha_{n}+\beta_{n})}{\bar{N}^{n}(\alpha_{1}+\beta_{1})^{n}}\frac{B^n}{(1-B)^{n}}\biggr)^{N-1}
\nonumber \\ && \times
\biggl(1+\sum_{n=1}^{\infty}\frac{(\alpha_{n}+\beta_{n})}{(\alpha_{1}+\beta_{1})^{n}}\frac{B^n}{(1-B)^{n}}\biggr)^{-N} \times \biggl[e^{[-NB_{n}-\bar{N}(1-B_{n})]}\biggr].
\end{eqnarray}
This is the general distribution function for any order of terms. Thus, we can write for $n=1$,  which is the case of weak interactions, and $n =2$ and $n = 3$,

\begin{eqnarray}
F(N)_1 &=& \frac{\bar{N}^{N}}{N!}\biggl(1+\frac{N}{\bar N}\frac{B}{(1-B)}\biggr)^{N-1}\biggl(1+\frac{B}{(1-B)}\biggr)^{-N} \times e^{(-NB-\bar N(1-B))},
 \nonumber \\
F(N)_2&=& \frac{\bar{N}^{N}}{N!}\biggl[1+\frac{NB}{\bar{N}(1-B)}+\frac{N^2(\alpha_{2}+\beta_{2})B^2}{\bar{N}^{2}(\alpha_{1}+\beta_{1})^2(1-B)^{2}}\biggr]^{N-1}\biggl[1+\frac{B}{(1-B)}+\frac{(\alpha_{2}+\beta_{2})B^2}{(\alpha_{1}+\beta_{1})^2(1-B)^{2}}\biggr]^{-N}\nonumber\\ && \times
 e^{-N\biggl(\frac{(\alpha_{1}+\beta_{1})^2B(1-B)+2(\alpha_{2}+\beta_{2})B^2}{(\alpha_{1}+\beta_{1})^{2}(1-B)^2+(\alpha_{1}+\beta_{1})^{2}B(1-B)+(\alpha_{2}+\beta_{2})B^2}\biggr)}
\times  e^{ -
\bar N\biggl(1-\frac{(\alpha_{1}+\beta_{1})^2B(1-B)+2(\alpha_{2}+\beta_{2})B^2}{(\alpha_{1}+\beta_{1})^{2}(1-B)^2+(\alpha_{1}+\beta_{1})^{2}B(1-B)+(\alpha_{2}+\beta_{2})B^2}\biggr)},
 \nonumber \\
F(N)_3 &=& \frac{\bar{N}^{N}}{N!}\biggl[1+\frac{NB}{\bar{N}(1-B)}+\frac{N^2(\alpha_{2}+\beta_{2})B^2}{\bar{N}^{2}(\alpha_{1}+\beta_{1})^2(1-B)^{2}}+\frac{N^3(\alpha_{3}+\beta_{3})B^3}{\bar{N}^{3}(\alpha_{1}+\beta_{1})^3(1-B)^{3}}\biggr]^{N-1}\nonumber\\
&&
\times \biggl[1+\frac{B}{(1-B)}+\frac{(\alpha_{2}+\beta_{2})B^2}{A_{1}^2(1-B)^{2}}+\frac{(\alpha_{3}+\beta_{3})B^3}{(\alpha_{1}+\beta_{1})^3(1-B)^{3}}\biggr]^{-N}
 \times e^{-N\biggl(1-\frac{(\alpha_{1}+\beta_{1})^2B(1-B)+2(\alpha_{2}+\beta_{2})B^2}{(\alpha_{1}+\beta_{1})(\alpha_{1}+\beta_{1})^{2}(1-B)^2+(\alpha_{1}+\beta_{1})^{2}B(1-B)+(\alpha_{2}+\beta_{2})B^2}\biggr)}
\nonumber  \\ && \times e^{ -
\bar N\biggl(1-\frac{(\alpha_{1}+\beta_{1})^2B(1-B)+2(\alpha_{2}+\beta_{2})B^2}{(\alpha_{1}+\beta_{1})^{2}(1-B)^2+(\alpha_{1}+\beta_{1})^{2}B(1-B)+(\alpha_{2}+\beta_{2})B^2}\biggr)}.\nonumber \\
\end{eqnarray}
Similarly, we can calculate the higher order distribution functions by taking higher order terms.
More the number of terms more clustering in the system.
Thus,  by including more terms we may have the distribution function of strongly interacting system of galaxies in the brane world models.
 \section{Comparative study}
We will be comparing our results with our previous work. As per the previous work the clustering parameter with first term is expressed as \cite{ahm14, ahm144}
\begin{equation}
b=\frac{\alpha_{1}x}{1+\alpha_{1}x}
\end{equation}
 Now after inclusion of brane corrections the clustering parameter can be expressed as
\begin{equation}
B=\frac{(\alpha_{1}+\beta_{1})x}{1+(\alpha_{1}+\beta_{1})x}
\end{equation}
The relation between $B$ and $b$ is obtained and we express it as
\begin{equation}
B=b\frac{1+\frac{\beta_{1}}{\alpha_{1}}}{1+\frac{\beta_{1}}{\alpha_{1}}b}
\end{equation}
where for $n=1$
\begin{equation}
a=\frac{k_l}{\epsilon}\frac{F(3/2,(l+1)/2;5/2;-R^{2}/\epsilon^{2})}{F(3/2,1/2;5/2;-R^{2}/\epsilon^{2})}
\end{equation}
where we have put
\begin{equation}
a=\frac{\beta_{1}}{\alpha_{1}}
\end{equation}
This ratio can be calculated for different values of $l$.
e.g for $l=1$ $a=0.29$, $l=2$ $a=0.103$, $l=3$ $a=0.046$, $l=4$ $a=0.025$ assuming $k_{l}=\epsilon$, $2\epsilon$
  We have   made a graphical comparison of distribution function $F(N)$ for different values of $l$ to see the effect of first brane term on clustering parameter. Furthermore,   $F(N)$ is also expressed in terms of $b$ as
\begin{eqnarray}
F(N)&=&\frac{\bar{N}^{N}}{N!}\biggl(1+\frac{N}{\bar N}\frac{\frac{(1+a)b}{1+ab}}{(1-\frac{(1+a)b}{1+ab})}\biggr)^{N-1}\biggl(1+\frac{\frac{(1+a)b}{1+ab}}{(1-\frac{(1+a)b}{1+ab})}\biggr)^{-N}
 \times e^{(-N\frac{(1+a)b}{1+ab}-\bar N(1-\frac{(1+a)b}{1+ab}))},
\end{eqnarray}
This is being compared with the earlier distribution function
\begin{eqnarray}
f(N)_1 &=& \frac{\bar{N}^{N}}{N!}\biggl(1+\frac{N}{\bar N}\frac{b}{(1-b)}\biggr)^{N-1}\biggl(1+\frac{b}{(1-b)}\biggr)^{-N}
 \times e^{(-Nb-\bar N(1-b))},
\end{eqnarray}\\

The graphs show the effect of Brane corrections on the distribution function.

\begin{figure}[hbp]
\centering
\begin{tabular}{cc}
\rotatebox{0}{\resizebox{70mm}{!}{\includegraphics{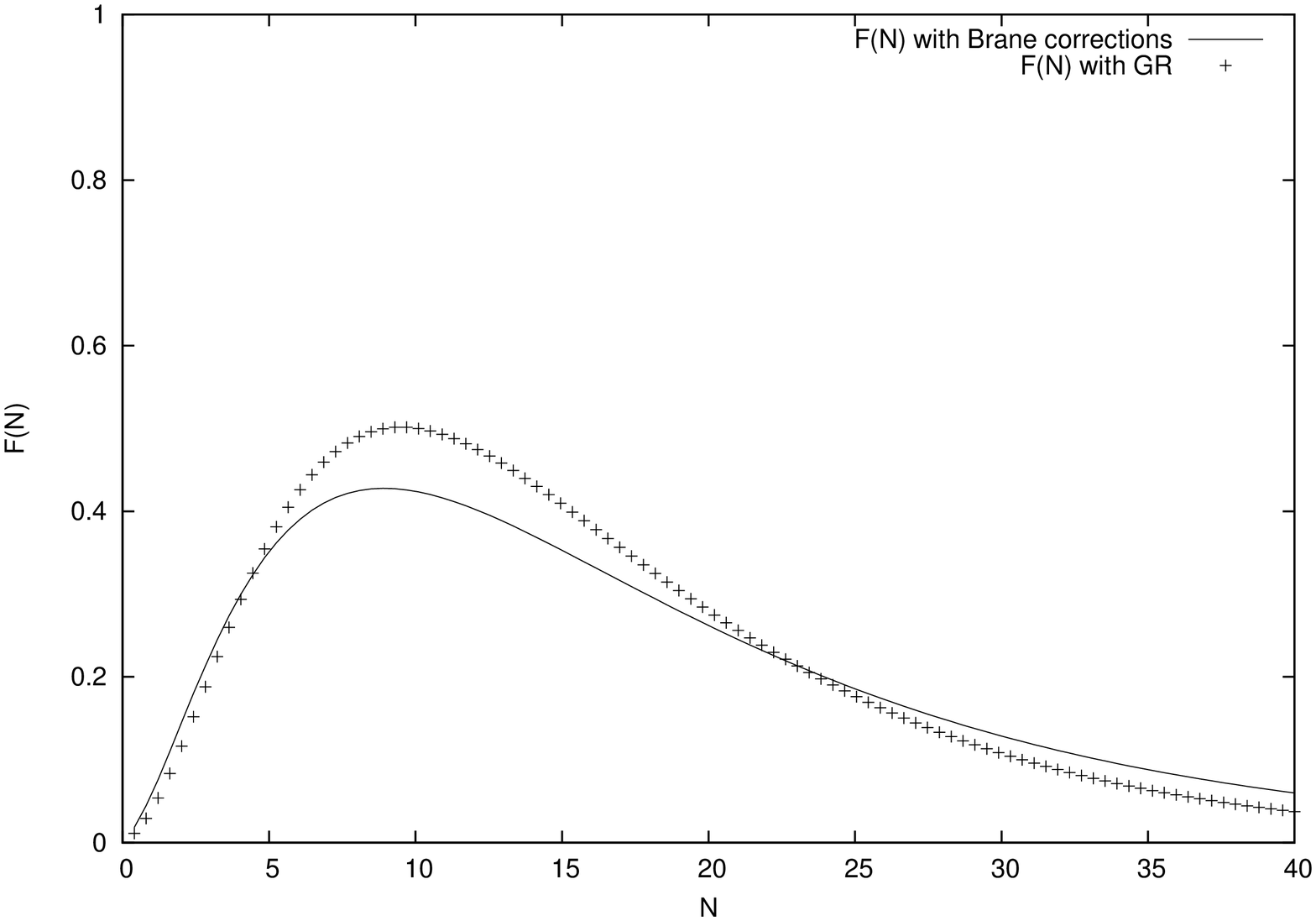}}}
\end{tabular}
\caption{Comparative study of distribution function $F(N)$ with brane correction and $f(N)$ without brane correction for $b=0.6$, $a=0.3$, $\bar{N}=10$ and $l=1$}
\end{figure}
\begin{figure}[hbp]
\centering
\begin{tabular}{cc}
\rotatebox{0}{\resizebox{70mm}{!}{\includegraphics{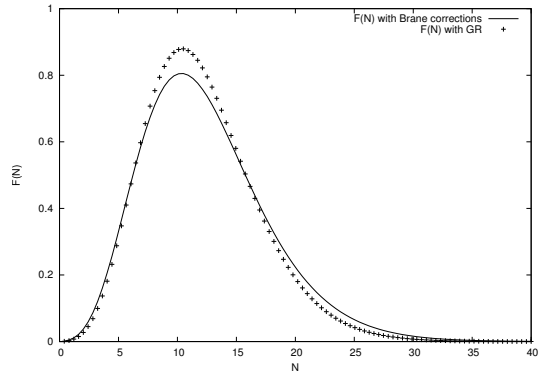}}}
\end{tabular}
\caption{Comparative study of distribution function $F(N)$ with brane correction and $f(N)$ without brane correction for $b=0.3$, $a=0.3$, $\bar{N}=10$ and $l=1$}
\end{figure}

\section{Two-Point Function}
In this section,  we will analyse the power law for galaxy-galaxy correlation function in Brane world.
It has been demonstrated that the
two-point   function $\xi(x)$ in clustering of galaxies obeys the following power law $r^{-1.8} $ \cite{pee80}. The
validity of this power law has been confirmed from N-body simulation \cite{sut90}.
We will now obtain a power law for the correlation function using Newtonian potential in brane world models. We will analyse the effect of the
the extended structure of galaxies and extra dimensions in brane world models on the power law. It will be demonstrated that both of these
generate higher order corrections in the form of the two-point correlation function.

In order to obtain the corrections to the power law from the extended structure of galaxies and extra dimensions in brane world models. let us
consider a spherical system of volume $V$ and energy
     $U$, such that it  contains   $N$ particles. Then we can write the following expression  \cite{sas00},
\begin{eqnarray}
U&=&\frac{3}{2}NT-\frac{N\rho}{2}\int_{V}\phi(r)\xi(r)4\pi r^2dr.
\end{eqnarray}
Using the interaction potential $\phi$ between two galaxies in brane world models
$
\phi= -\frac{Gm^2}{(r^2+\epsilon^2)^{1/2}}\biggl(1+\frac{k_l}{(r^2+\epsilon^2)^{l/2}}\biggr),
$
and this expression for the energy of a spherical system (for $n=1$), we obtain
\begin{eqnarray}
B&=&\frac{Gm^2\rho}{6T}\int\frac{\xi(r)}{r}(1+\frac{\epsilon^2}{r^2})^{-1/2}(1+k_l(r^2+\epsilon^2)^{-l/2})dV.
\end{eqnarray}
Now we also using
\begin{eqnarray}
\frac{\partial B}{\partial V}&=&\frac{Gm^2\rho}{6T}\frac{\partial}{\partial V}\int\frac{\xi(r)}{r}(1+\frac{\epsilon^2}{r^2})^{-1/2}
(1+k_l(r^2+\epsilon^2)^{-l/2})dV \nonumber\\  && +\frac{Gm^2\rho}{6T}\int\frac{\xi(r)}{r}(1+\frac{\epsilon^2}{r^2})^{-1/2}
(1+k_l(r^2+\epsilon^2)^{-l/2})dV\frac{1}{\rho}\frac{d\rho}{dV},
\nonumber \\
\frac{\partial B}{\partial V}&=&\frac{\partial \rho}{\partial V}\frac{\partial B}{\partial \rho},
\end{eqnarray} we obtain
\begin{eqnarray}
\frac{\partial B}{\partial \rho}&=&\frac{B(1-B)}{\rho}.
\end{eqnarray}
Furthermore, using  $\rho=\frac{N}{V}$, we obtain
\begin{eqnarray}
\frac{\partial \rho}{\partial V}&=&-\frac{\rho}{V}
\end{eqnarray}
Neglecting higher powers of $\epsilon/r$, we obtain the following power law
\begin{eqnarray}
\xi(r)&=&\frac{9B^2T}{2\pi Gm^2\rho}\frac{1}{r^2}\bigl(1+\frac{\epsilon^2}{2r^2}-\frac{k_l}{r^l}\bigr)
\end{eqnarray}
If we neglect both these effects, we observe that we the power law is very similar
  Peebles power law $r^{-1.8} \sim r^{-2}$. However, this law receives higher order corrections by taking into account the fact
  that galaxies have an extended structure and by assuming the existence of extra dimensions.
The term with $\epsilon $ measures the effect of having an extended structure, and
the term with $k_l$ measures the effect of have extra dimensions in brane world models. Both these effects have
opposite effects on the power law. This it is possible to use this behavior of two-point correlation function
to test astrophysical data. In case, the astrophysical data shows a positive deviation from the power law, it will
support the idea that the extended structure of galaxies can not be neglected in analysing the clustering of galaxies.
However, a negative deviation from the power law will support the existence of extra dimensions in the brane world models.

\section{Conclusion}
In this paper,   we have analyzed the clustering of galaxies using modified Newtonian potential. The modification we considered was motivated by the brane world models.
It is well known that in brane world models, the standard model fields are fixed on the brane, but the gravitons propagate freely into the bulk.
This changes the effective gravitational potential, and the deviation from the Newtonian potential can be observed at large distances. Thus,
it is both interesting and important to use the  brane world modified Newtonian potential for analyzing the clustering of galaxies.
So, in this paper, we analyze the effects of extra dimensions on the clustering of galaxies in the brane world models.
We calculate the partition functions for a system of galaxies which interact through the modified Newtonian potential.
This partition function is then used for calculating  important the thermodynamic quantities of this system. A
 general clustering parameter is also obtained for      galaxies interacting in a brane world model. Finally, we also study the effect of the modified Newtonian potential on the two-point function between galaxies.

 We have used the power law deformation of the Newtonian potential which occurs in the Randall-Sundrum model where extra dimensions are allowed to be
 infinite. In other brane world models, the modification of the Newtonian potential are parametrized by a Yukawa potential. It would be interesting
 to repeat this analysis for the Newtonian potential  modified by a Yukawa potential.
 It may be noted that the eqnarrays of state for the galaxies interacting through the usual Newtonian potential
has been  obtained and discussed for point mass galaxies
\cite{sas84, ahm02}. These derivations were based on
fundamental thermodynamic principles \cite{sas84,  sas00}. Analytic expression for the grand canonical partition functions of point masses
has also been analyzed \cite{ahm02}. Subsequently, these were generalized to extended masses
i.e. galaxies with halos, which cluster gravitationally in an
expanding universe. From the partition functions, the thermodynamic properties, distribution functions, void distribution functions and moments of
distributions have been obtained more rigorously from statistical mechanical basis \cite{ahm02}.   It would be interesting to obtain these results for
different brane world models and compare the results with observations. It will be interesting to investigate the other forms of modified Newton's law of gravity in other theories like f(R), entropic force gravity and noncommutative geometry. We hope to report on these in the future.

\section{Appendix}
We write the calculations of the first integral of equation (17), the second integral has similar calculations.
\begin{eqnarray}
I_{1}(T,V)&=&4\pi V\biggl[\big(\frac{Gm^{2}}{T}\big)^{0}\int_{0}^{R_{1}}r^{2}dr+\frac{1}{1!}\big(\frac{Gm^{2}}{T}\big)^{1}\int_{0}^{R_{1}}\frac{r^{2}}{(r^{2}+
\epsilon^{2})^{1/2}}\nonumber\\&+&\frac{1}{2!}\big(\frac{Gm^{2}}{T}\big)^{2}\int_{0}^{R_{1}}\frac{r^{2}}{(r^{2}+\epsilon^{2})}dr+
\frac{1}{3!}\big(\frac{Gm^{2}}{T}\big)^{3}\int_{0}^{R_{1}}\frac{r^{2}}{(r^{2}+\epsilon^{2})^{3/2}}dr\nonumber\\
&+&\frac{1}{4!}\big(\frac{Gm^{2}}{T}\big)^{4}\int_{0}^{R_{1}}\frac{r^{2}}{(r^{2}+\epsilon^{2})^{2}}dr+
\frac{1}{5!}\big(\frac{Gm^{2}}{T}\big)^{5}\int_{0}^{R_{1}}\frac{r^{2}}{(r^{2}+\epsilon^{2})^{5/2}}dr\nonumber\\
&+&\frac{1}{6!}\big(\frac{Gm^{2}}{T}\big)^{6}\int_{0}^{R_{1}}\frac{r^{2}}{(r^{2}+\epsilon^{2})^{3}}dr+
\frac{1}{7!}\big(\frac{Gm^{2}}{T}\big)^{7}\int_{0}^{R_{1}}\frac{r^{2}}{(r^{2}+\epsilon^{2})^{7/2}}dr\nonumber\\
&+&\frac{1}{8!}\big(\frac{Gm^{2}}{T}\big)^{8}\int_{0}^{R_{1}}\frac{r^{2}}{(r^{2}+\epsilon^{2})^{4}}dr+\dots\biggr]
\end{eqnarray}

\begin{eqnarray}
I_{1}(T,V)&=&4\pi V\biggl[\frac{R_{1}^{3}}{3}+\biggl(\frac{Gm^{2}}{T}\biggr)\biggl(\frac{R_{1}^{2}}{2}\sqrt{1+\frac{\epsilon^2}{R_{1}^{2}}}
+\frac{\epsilon^{2}}{2}ln\frac{\frac{\epsilon}{R_{1}}}{1+\sqrt{1+\frac{\epsilon^2}{R_{1}^{2}}}}\biggr)\nonumber\\
&+&\frac{1}{2!}\biggl(\frac{Gm^2}{T}\biggr)^2\biggl(R_{1}-\epsilon tan^{-1}(\frac{1}{\epsilon/R_{1}})\biggr)+\frac{1}{3!}\biggl(\frac{Gm^2}{T}\biggr)^3\nonumber\\
&& \biggl(-\frac{1}{\sqrt{1+\frac{\epsilon^2}{R_{1}^{2}}}}-ln\frac{\frac{\epsilon}{R_{1}}}{1+\sqrt{1+\frac{\epsilon^2}{R_{1}^{2}}}}\biggr)
+\frac{1}{4!}\biggl(\frac{Gm^{2}}{T}\biggr)^4\nonumber\\ && \biggl(-\frac{1}{2R_{1}(1+\frac{\epsilon^2}{R_{1}^2})}
+\frac{1}{2\epsilon}tan^{-1}\frac{1}{\epsilon/R_{1}}\biggr)+\frac{1}{5!}\biggl(\frac{Gm^2}{T}\biggr)^5\nonumber\\
&&\biggl(\frac{1}{3R_{1}^{3}(1+\frac{\epsilon^2}{R_{1}^2})^3}\biggr)
+\frac{1}{6!}\biggl(\frac{Gm^2}{T}\biggr)^6\biggl(-\frac{1}{4R_{1}^3(1+\frac{\epsilon^2}{R_{1}^2})^2}
+\frac{1}{\epsilon^2R_{1}(1+\frac{\epsilon^2}{R_{1}^2})}\nonumber\\
&+&\frac{1}{8\epsilon^3}tan^{-1}(\frac{1}{\epsilon/R_{1}})\biggr)
+\frac{1}{7!}\biggl(\frac{Gm^2}{T}\biggr)^7
\biggl(\frac{1}{3\epsilon^4(1+\frac{\epsilon^2}{R_{1}^2})^{3/2}}-
\frac{1}{5\epsilon^4(1+\frac{\epsilon^2}{R_{1}^2})^{5/2}}\biggr)+\dots\biggr]
\end{eqnarray}
\newpage
\begin{eqnarray}
I_{1}(T,V)&=&4\pi V\biggl[\frac{R_{1}^{3}}{3}+\biggl(\frac{3Gm^{2}}{2TR_1}\biggr)\biggl(\frac{2R_1}{3\epsilon}\biggr)\biggl(\frac{R_1^{3}}{3}\frac{3\epsilon}{R_1^{3}}\biggr)\nonumber\\
&&\biggl(\frac{R_{1}^{2}}{2}\sqrt{1+\frac{\epsilon^2}{R_{1}^{2}}}+ \frac{\epsilon^{2}}{2}ln\frac{\frac{\epsilon}{R_{1}}}{1+\sqrt{1+\frac{\epsilon^2}{R_{1}^{2}}}}\biggr)
+\frac{1}{2!}\biggl(\frac{3Gm^2}{2TR_1}\biggr)^2\nonumber\\
&&\biggl(\frac{2R_1}{3\epsilon}\biggr)^2\biggl(\frac{R_1^{3}}{3}\frac{3\epsilon^2}{R_1^{3}}\biggr)
\biggl(R_{1}-\epsilon tan^{-1}(\frac{1}{\epsilon/R_{1}})\biggr)
+\frac{1}{3!}\biggl(\frac{3Gm^2}{2TR_1}\biggr)^3\nonumber\\
&&\biggl(\frac{2R_1}{3\epsilon}\biggr)^3\biggl(\frac{R_1^{3}}{3}\frac{3\epsilon^3}{R_1^{3}}\biggr)\biggl(-\frac{1}{\sqrt{1+\frac{\epsilon^2}{R_{1}^{2}}}}-ln\frac{\frac{\epsilon}{R_{1}}}{1+\sqrt{1+\frac{\epsilon^2}{R_{1}^{2}}}}\biggr)
+\frac{1}{4!}\biggl(\frac{3Gm^{2}}{2TR_1}\biggr)^4\nonumber\\
&&\biggl(\frac{2R_1}{3\epsilon}\biggr)^4\biggl(\frac{R_1^{3}}{3}\frac{3\epsilon^4}{R_1^{3}}\biggr)\biggl(-\frac{1}{2R_{1}(1+\frac{\epsilon^2}{R_{1}^2})}+\frac{1}{2\epsilon}tan^{-1}\frac{1}{\epsilon/R_{1}}\biggr)\nonumber\\
&+&\frac{1}{5!}\biggl(\frac{3Gm^2}{2TR_1}\biggr)^5\biggl(\frac{2R_1}{3\epsilon}\biggr)^5\biggl(\frac{R_1^{3}}{3}\frac{3\epsilon^5}{R_1^{3}}\biggr) \biggl(\frac{1}{3R_{1}^{3}(1+\frac{\epsilon^2}{R_{1}^2})^3}\biggr)
+\frac{1}{6!}\biggl(\frac{3Gm^2}{2TR_1}\biggr)^6\nonumber\\
&&\biggl(\frac{2R_1}{3\epsilon}\biggr)^6\biggl(\frac{R_1^{3}}{3}\frac{3\epsilon^6}{R_1^{3}}\biggr)\biggl(-\frac{1}{4R_{1}^3(1+\frac{\epsilon^2}{R_{1}^2})^2}+\frac{1}{\epsilon^2R_{1}(1+\frac{\epsilon^2}{R_{1}^2})}+\frac{1}{8\epsilon^3}tan^{-1}(\frac{1}{\epsilon/R_{1}})\biggr)+\dots\biggr]
\end{eqnarray}
\begin{eqnarray}
I_{1}(T,V)=V^2\biggl(1+x\biggl(\frac{1}{1!}\biggl(\frac{2R_{1}}{3\epsilon}\biggr)^{1} F \biggl(\frac{3}{2},\frac{1}{2};\frac{5}{2};-\frac{R_{1}^2}{\epsilon^2}\biggr)
+x^2\frac{1}{2!}\biggl(\frac{2R_{1}}{3\epsilon}\biggr)^{2} F \biggl(\frac{3}{2},\frac{2}{2};\frac{5}{2};-\frac{R_{1}^2}{\epsilon^2}\biggr)\nonumber\\
+x^3\frac{1}{3!}\biggl(\frac{2R_{1}}{3\epsilon}\biggr)^{3} F \biggl(\frac{3}{2},\frac{3}{2};\frac{5}{2};-\frac{R_{1}^2}{\epsilon^2}\biggr)
+x^4\frac{1}{4!}\biggl(\frac{2R_{1}}{3\epsilon}\biggr)^{4} F \biggl(\frac{3}{2},\frac{4}{2};\frac{5}{2};-\frac{R_{1}^2}{\epsilon^2}\biggr)\nonumber\\+x^5\frac{1}{5!}\biggl(\frac{2R_{1}}{3\epsilon}\biggr)^{5} F \biggl(\frac{3}{2},\frac{5}{2};\frac{5}{2};-\frac{R_{1}^2}{\epsilon^2}\biggr)+x^6\frac{1}{6!}\biggl(\frac{2R_{1}}{3\epsilon}\biggr)^{6} F \biggl(\frac{3}{2},\frac{6}{2};\frac{5}{2};-\frac{R_{1}^2}{\epsilon^2}\biggr)+\dots \biggr)
\end{eqnarray}
where
\begin{eqnarray}
F \biggl(\frac{3}{2},\frac{1}{2};\frac{5}{2};-\frac{R_{1}^2}{\epsilon^2}\biggr)=\frac{3\epsilon}{R_1^3}\biggl(\frac{R_{1}^{2}}{2}\sqrt{1+\frac{\epsilon^2}{R_{1}^{2}}}+\frac{\epsilon^{2}}{2}ln\frac{\frac{\epsilon}{R_{1}}}{1+\sqrt{1+\frac{\epsilon^2}{R_{1}^{2}}}}\biggr)
\end{eqnarray}
\begin{eqnarray}
F \biggl(\frac{3}{2},\frac{2}{2};\frac{5}{2};-\frac{R_{1}^2}{\epsilon^2}\biggr)=\frac{3\epsilon^2}{R_1^3}\biggl(R_{1}-\epsilon tan^{-1}(\frac{1}{\epsilon/R_{1}})\biggr)
\end{eqnarray}
\begin{eqnarray}
F \biggl(\frac{3}{2},\frac{3}{2};\frac{5}{2};-\frac{R_{1}^2}{\epsilon^2}\biggr)=\frac{3\epsilon^3}{R_1^3}\biggl(-\frac{1}{\sqrt{1+\frac{\epsilon^2}{R_{1}^{2}}}}-ln\frac{\frac{\epsilon}{R_{1}}}{1+\sqrt{1+\frac{\epsilon^2}{R_{1}^{2}}}}\biggr)
\end{eqnarray}
\begin{eqnarray}
F \biggl(\frac{3}{2},\frac{4}{2};\frac{5}{2};-\frac{R_{1}^2}{\epsilon^2}\biggr)=\frac{3\epsilon^4}{R_1^3}\biggl(-\frac{1}{2R_{1}(1+\frac{\epsilon^2}{R_{1}^2})}+\frac{1}{2\epsilon}tan^{-1}\frac{1}{\epsilon/R_{1}}\biggr)
\end{eqnarray}
\begin{eqnarray}
F \biggl(\frac{3}{2},\frac{5}{2};\frac{5}{2};-\frac{R_{1}^2}{\epsilon^2}\biggr)=\frac{3\epsilon^5}{R_1^3}\biggl(\frac{1}{3R_{1}^{3}(1+\frac{\epsilon^2}{R_{1}^2})^3}\biggr)
\end{eqnarray}
\begin{eqnarray}
F \biggl(\frac{3}{2},\frac{6}{2};\frac{5}{2};-\frac{R_{1}^2}{\epsilon^2}\biggr)=\frac{3\epsilon^6}{R_1^3}\biggl(-\frac{1}{4R_{1}^3(1+\frac{\epsilon^2}{R_{1}^2})^2}+\frac{1}{\epsilon^2R_{1}(1+\frac{\epsilon^2}{R_{1}^2})}\nonumber\\+\frac{1}{8\epsilon^3}tan^{-1}(\frac{1}{\epsilon/R_{1}})\biggr)
\end{eqnarray}
In general
\begin{eqnarray}
I_{1}(T,V)=V^2\biggl(1+\sum_{n=1}^\infty x^n\frac{1}{n!}\biggl(\frac{2R_{1}}{3\epsilon}\biggr)^{n} F \biggl(\frac{3}{2},\frac{n}{2};\frac{5}{2};-\frac{R_{1}^2}{\epsilon^2}\biggr)\biggr)
\end{eqnarray}
\end{document}